\documentclass[9pt,conference]{IEEEtran}

\usepackage{verbatim}
\usepackage{color}
\usepackage{amsthm}
\usepackage{amsmath}
\usepackage{amssymb}
\usepackage{graphicx}
\usepackage{mathrsfs}
\usepackage{mdwlist}
\usepackage{amsthm}
\usepackage[ruled]{algorithm2e}
\usepackage{url}
\usepackage{cite}
\usepackage[latin9]{inputenc}
\usepackage[active]{srcltx}
\IEEEoverridecommandlockouts
\newtheorem{theorem}{Theorem}
\newtheorem{proposition}{Proposition}

\newtheorem{remark}{Remark}
\usepackage[footnotesize]{caption}

\DeclareMathOperator*{\argmin}{argmin}
\newcommand{\norm}[1]{\left\lVert#1\right\rVert}

\makeatletter
\def\ps@IEEEtitlepagestyle{%
	\def\@oddfoot{\mycopyrightnotice}%
	\def\@evenfoot{}%
}
\def\mycopyrightnotice{%
	{\footnotesize Copyright 2001 SS\&C. Published in the Proceedings of the 50th annual Asilomar conference on signals, systems, and computers, Nov. 6-9, 2016, CA, USA.\hfill}% <--- Change here
	\gdef\mycopyrightnotice{}% just in case
}

\begin{document}

\title{\vspace{-0.7cm}Distributed Nonconvex Multiagent Optimization Over Time-Varying Networks\vspace{-0.2cm}}

\author{Ying Sun, Gesualdo Scutari, and Daniel Palomar$^\dagger$\vspace{-0.9cm}
\thanks{$^\dagger$Sun and Scutari are with   the School of Industrial Engineering, Purdue University, West-Lafayette, IN, USA; emails: \texttt{<sun578,gscutari>@purdue.edu}. Palomar is with the Hong Kong University of Science and Technology (HKUST), Hong Kong; email: \texttt{palomar@ust.hk}.   The work of  Ying and Scutari was supported by the USA National Science Foundation under Grants CIF 1564044 and CAREER Award 1555850, and the ONR N00014-16-1-2244. The work of Palomar was supported by the Hong Kong RGC 16207814 research grant.}}

 \maketitle

\begin{abstract}
We study \emph{nonconvex} distributed optimization in multiagent networks where the communications between nodes is modeled as a time-varying sequence of \emph{arbitrary} digraphs. We introduce a novel broadcast-based distributed algorithmic framework for the (constrained)  minimization of the sum of a smooth (possibly nonconvex and nonseparable) function, \textit{i.e.}, the agents' sum-utility, plus a convex (possibly nonsmooth and nonseparable) regularizer. The latter is usually employed to enforce some structure in the solution, typically sparsity. The proposed method hinges on  Successive Convex Approximation (SCA) techniques
coupled with i) a  tracking mechanism instrumental
to locally estimate the gradients of agents' cost functions; and ii) a
 novel broadcast protocol to disseminate information and distribute  the computation  
among the agents. %successive convex approximation techniques and a novel broadcast protocol to disseminate information and distribute  the computation over the network.  
 Asymptotic convergence to stationary solutions is established. A key feature of the proposed algorithm is that it neither requires the double-stochasticity of the consensus matrices (but only column stochasticity) nor the knowledge of the graph sequence to implement.
 %Moreover, it requires no knowledge of the graph sequence to implement.  
 To the best of our knowledge, the proposed framework is the first broadcast-based  distributed algorithm for convex and \emph{nonconvex constrained} optimization  over \emph{arbitrary, time-varying} digraphs. Numerical results show that our  algorithm outperforms current schemes on both convex and nonconvex problems.\vspace{-0.1cm}
\end{abstract}
 
\section{Introduction}
Distributed optimization has found  wide range of applications in several areas, including   %received significant attention due to the growing interest in solving  large-scale optimizaiton problems in 
machine learning, data analysis, signal processing, networking, and decentralized control. %for wireless communication networks, power networks, and sensor networks, just to name a few. 
Common to these problems is a network of agents%A central generic problem in such applications is  distributed optimization in
%multiagent networks wherein  agents$-$which may be 
$-$processors, computers of a cluster, nodes of a sensor network, vehicles, or UAVs$-$that want to cooperatively minimize  a global cost function  by means of actions taken by each agent and local coordination between neighboring nodes. In this paper,  we consider the following general class of (possibly)  \emph{nonconvex}  multiagent problems:\vspace{-0.1cm}
%Many in-network optimization problems can be formulated as optimizing the sum of private objective functions locally belong to each agent. This general formulation enjoys a wide range of applications from areas including data analysis, statistical estimation, communication, as well as machine learning. Centralized optimization algorithms require each agent knowing the objective function before optimization, which %is impossible in general since it asks for the transmission of functions, which is infinite-dimensional, among agents. Even in special cases that the functions can be represented by finite-dimensional parameters, it
%can  incur a prohibitively large amount of communication cost. Besides the problem of implementation feasibility, agents may unwilling to share their local utility function due to security reasons. Consequently, developing  distributed  optimization  algorithms has attracted a significant attention.
%We consider a network composed of $I$ agents collaboratively solving a sum-utility optimization problem of the form
\begin{equation}
\begin{aligned}
\min_{\bf{x}\in \mathcal K}&&U\left(\bf x\right)\triangleq \sum_{i=1}^I f_i\left(\bf x\right) + G\left(\bf x\right),
\end{aligned}\label{eq: P}\vspace{-0.1cm}
\end{equation}
where  $f_i:\mathbb{R}^m\to \mathbb{R}$ is the cost function of agent $i$, assumed to be   smooth but (possibly) nonconvex;  $G:\mathbb{R}^m\to \mathbb{R}$ is a convex (possibly nonsmooth) regularizer;  and $\mathcal{K}$ is a closed convex subset of $\mathbb{R}^m$. Usually the nonsmooth term
is used to promote some extra structure in the solution; for
instance,    $G({\bf x})=c\,\|{\bf x}\|_1$ or $G({\bf x
})=c \,\sum_{i=1}^N\|{\bf x}_i\|_2$ are widely used to impose (group)  sparsity of the  solution. 
This general formulation arises naturally from many applications, including  statistical inference over (e.g., sensor and power) networks, formation control,  spectrum access coordination,  distributed machine learning (e.g., LASSO, logistic regression, dictionary learning, matrix completion, tensor factorization), resource allocation problems  in wireless communication networks, and distributed ``epidemic'' message routing in networks.%Common to these problems is the necessity of performing  a completely decentralized  computation/optimization.

Our goal  is  developing solution methods for the \emph{nonconvex} problem (\ref{eq: P}) in the following distributed setting: i) Each agent $i$ knows only its own function  $f_i$ (as well as $G$ and $\mathcal K$); and ii) the communication topology connecting the agents  is \emph{time-varying} and \emph{directed}, and it is not known to the agents.  Time-varying communication topologies arise, for instance, in mobile  wireless networks, wherein the nodes are mobile and/or communicate throughout  (fast-)fading channels. Directed communication links are also a natural assumption as in many cases there is no reason to expect different nodes to transmit at the same power level or that transmitter and receivers are geographically collocated (e.g., think of ad-hoc networks).  
 
Distributed solution methods  for \emph{convex} instances of Problem  (\ref{eq: P}) have been widely studied in the literature, under various assumptions on network topology; some recent contributions include  \cite{nedic2009distributed,shi2015extra,srivastava2011distributed,duchi2012dual,gharesifard2014distributed,
tsianos2012push,tsianos2011distributed,nedic2010constrained}. The majority of the aforementioned  works  assume either  undirected  graphs or  static directed graphs. Moreover all the algorithms developed in the aforementioned papers  along with their convergence
analysis are not applicable to nonconvex problems, and thus to  Problem (\ref{eq: P}). %algorithms and their convergence analyses propsoed in the aforementioned paper are not of them are   In  the noncovex case, results are instead  scarce; 
We are aware of only   few works dealing with distributed algorithms for some nonconvex instances of   (\ref{eq: P}), namely: \cite{zhu2013approximate,bianchi2011convergence,Lorenzo2016NEXT,
tatarenko2015non}. %In \cite{zhu2013approximate}, a consensus-based distributed dual-subgradient
%algorithm was studied over static undirected graphs.The method calls for the global solution
%of possibly difficult nonconvex subproblems, and it does not findtatarenko2015non
%(stationary) solutions of the original problem but those of an auxiliary
%problem, which are not necessarily stationary for the original
%problem. In \cite{bianchi2011convergence}, the authors studied convergence of a distributed
%stochastic projection algorithm involving random gossip between
%agents. However, the scheme is not applicable to Problem (\ref{eq: P}) when
%$G=0$. The nonconvex optimization problem in the general form
%(\ref{eq: P}) was tackled for the first time in our recent work \cite{Lorenzo2016NEXT}. However, the algorithms proposed in these papers requires specific network topology   and full knowledge of the graph sequence to implement. In fact,  some form of balancedness in the graph is needed, often reflected in a requirement of having a sequence of doubly stochastic matrices that are commensurate with the sequence of underlying communication graphs. 
  Among them,  our previous work \cite{Lorenzo2016NEXT} is to date the only method applicable to the general class of \emph{nonconvex constrained} problems in the form    (\ref{eq: P}). However, the implementability of   algorithm   \cite{Lorenzo2016NEXT}  relies on the possibility of building  a sequence of \emph{double-stochastic} consensus matrices that are commensurate with the sequence of underlying time-varying communication digraphs.  %(i.e., the element $(i,j)$ of the  consensus matrix at a given iteration is the weight given by agent $i$  to the message received from  $j$ at the current iteration, and is equal to zero in case agent $i$  receives no message from  $j$). 
This  can  limit  the applicability of the method in practice, especially when the network topology is  time-varying, for several reasons. First, not all digraphs are  doubly-stochasticable (i.e., admit   a doubly stochastic   adjacency matrix); some form of balancedness in the graph is needed \cite{gharesifard2010does}, which limits the class of network topologies over which algorithm   \cite{Lorenzo2016NEXT} can be applied. Moreover, necessary and sufficient conditions for a digraph to be doubly-stochasticable are not easy to be checked in practice. Second, constructing a doubly-stochastic weight matrix matching the graph, even when possible, calls for   computationally intense, generally centralized,  algorithms.  Third, %to implement the algorithm  in \cite{Lorenzo2016NEXT}, agents need to know the sequence of the digraphs.  Fourth, 
double stochasticity prevents one  from using natural broadcast schemes, in which a given agent may transmit its local estimate to all its neighbors without expecting any immediate feedback. 

%To summarize, the above analysis shows that current distributed algorithms have the  the following two limitations: 1) they are    not applicable to nonconvex objective functions; and 2)  algorithms for time-varying digraphs either assume specific graph topology (doubly stochaticable), or require the problem to be unconstrained.
The analysis of the literature shows that the design of distributed algorithms for the class of problems (\ref{eq: P}) over \emph{time-varying, arbitrary} digraphs is up to date a challenging and open problem, \emph{even in the case of convex cost functions} $f_i$. This paper introduces the first \emph{broadcast}-based distributed algorithmic framework for the aforementioned class of problems.  
 The crux of the framework is a general
convexification-decomposition technique that hinges on our
recent (primal) Successive Convex Approximation (SCA)
methods  \cite{ScuFacSonPalPan2014,facchinei2015parallel}, coupled with i) a tacking mechanism that allows every agent $i$ to estimate locally the gradients of other agents' functions $\sum_{j\neq i}f_j$; and ii) a novel broadcast  protocol instrumental  to distribute the computation and propagate the needed information over the network. We term the new scheme ``distributed Successive cONvex Approximation algorithm over Time-varying digrAphs (SONATA)''. Some key desirable features of SONATA are: i) It is  applicable to \emph{arbitrary} (possibly) time-varying network topologies; ii) it is \emph{fully distributed},  requiring neither the knowledge  of the graph sequence nor the use of a double-stochastic consensus matrix; in fact, each agent  just needs to broadcast its local estimates to all its neighbors  without expecting any feedback; iii) it deals with nonconvex and nonsmooth
objectives as well as (convex) constraints; and iv) it is  
very flexible in the choice of the approximations of $f_i$'s, which
need not be necessarily its first or second order approximation
(like in all current distributed gradient schemes). Asymptotic convergence to stationary solutions of Problem (\ref{eq: P}) is proved.  
Numerical results show that %the customization of 
SONATA, applied to a number of \emph{convex} and \emph{nonconvex}  problems, %applications over time-varying networks 
outperforms state-of-the-art schemes, in terms of practical convergence while reaching the
same (stationary) solutions. As a final remark,  we point out   that 
the proposed  broadcast protocol  is different from the renowned pushed-sum protocol  \cite{kempe2003gossip}, used in a number of papers    \cite{tsianos2012push,tsianos2011distributed,nedic2015distributed,tatarenko2015non}  to remove the double-stochastic requirement on the consensus matrix. The major difference is that the proposed method is the first one applicable to \emph{constrained} (convex and nonconvex) optimization problems while push-sum-based schemes  work only for unconstrained  problems (this is because  push-sum-based  updates do not
preserve feasibility of the iterates).

The rest of the paper is  organized  as follows. In Section \ref{sec: Algorithm} we first introduce the general idea of SONATA, followed by its formal description  along with  its convergence properties.  Section \ref{sec:connection} sheds light on  the connection between  SONATA and  some recent  distributed algorithms proposed in the literature (mostly appeared after the submission of this work). Some applications of SONATA are discussed  in Section \ref{sec: Numerical} along with some numerical results. Finally,  Section \ref{sec: conclusion} draws some conclusions.\vspace{-0.2cm}
\\
%2. problem formulation
%3. a review on methods in the convex case, including admm, subgradient-push (Nedich)
%4. state the limitations:  convexity, unconstrained/constrained with doubly stochaticable digraph
%5. our contribution: nonconvex optimization with time-varying graph and constraints
%6. a review of the literature on non-convex cases
%    6.1 two works without convergence
%    6.2 Touri's paper, smooth objective function, gradient method, without constraint
%	6.3 NEXT, graph topology limitation
%7. organization
%8. notation
	
\section{Algorithmic Design}\label{sec: Algorithm}
%In this section, we first introduce technical assumptions on the objective function, followed by a mathematical characterization of the time-varying digraphs we are interested in. The push-sum consensus protocol is presented subsequently. In the end, we give an informal description of SONATA to illustrate the main idea.
%\subsection{Preliminaries}
We study  Problem \eqref{eq: P} under the following standard assumptions. 

\noindent\textbf{Assumption A (Problem Setup)}
\begin{enumerate*}
	\item[\textbf{(A1)}] The set $\mathcal{K}\neq\emptyset$ is closed and convex;
	\item[\textbf{(A2)}] Each $f_i$ is a continuously differentiable function defined on an open set containing $\mathcal{K}$;
	\item[\textbf{(A3)}] Each $\nabla f_i$  is Lipschitz continuous on $\mathcal{K}$; %with Lipschitz constant $L_i$;
	\item[\textbf{(A4)}] $\nabla F$ is bounded on $\mathcal{K}$, with $F(\mathbf{x})=\sum_if_i(\mathbf{x})$; %, \textit{i.e.}, there exists a constant $L_F\in \left(0,+\infty\right)$ such that $\|\nabla F\left(\bf x\right)\|\leq F_L,\,\forall \mathbf{x}\in \mathcal{K}$;
	\item[\textbf{(A5)}]  $G$ is convex with bounded subgradients on $\mathcal{K}$;%, \textit{i.e.}, there exists a constant $L_G\in \left(0,+\infty\right)$ such that $\|\partial G\left(\bf x\right)\|\leq F_G,\,\forall \mathbf{x}\in \mathcal{K}$;
	\item[\textbf{(A6)}] $U$ is coercive on $\mathcal K$, i.e., $\lim_{\mathbf{x}\in\mathcal{K},\,\|\mathbf{x}\|\to \infty} U\left(\mathbf{x}\right)= +\infty$.
\end{enumerate*}\smallskip

Assumption A is standard and satisfied by many practical
problems. For instance, A3-A5 hold automatically if $\mathcal K$ is
bounded and $f_i$ is twice continuously differentiable, whereas A6 guarantees the existence of a solution.
Note that each $f_i$ need not be convex and  is known only  by agent $i$.
%. Under this set-up, the agents need to communicate with each other in order to solve Problem \eqref{eq: P}.

\noindent\textbf{On the network topology.} Time is slotted and, at each time-slot $n$, the network of agents is  modeled as a  time-varying digraph  $\mathcal{G}\left[n\right]=\left(\mathcal{V},\mathcal{E}\left[n\right]\right))$, where the set of vertices $\mathcal{V}=\{1,\ldots,I\}$ represents the $I$ agents, and the set of edges $\mathcal{E}\left[n\right]$ represents the agents' communication links. The  in-neighborhood of agent $i$ at time $n$ (including node $i$) is defined as  $\mathcal{N}_i^{\rm in}[n]=\{j|(j,i)\in\mathcal{E}[n]\}\cup\{i\}$ whereas its out-neighbor is defined as $\mathcal{N}_i^\textrm{out}\left[n\right]=\{j|\left(i,j\right)\in\mathcal{E}\left[n\right]\}\cup\{i\}$.
Agent $i$ can receive information from its in-neighbors, and  send information to its out neighbors. The out-degree of agent $i$ is defined as $d_i\left[n\right] \triangleq  \left|\mathcal{N}_i^\textrm{out}\left[n\right]\right|$.  To let information propagate over the network,  we assume that  the graph sequence $\left(\mathcal{G}\left[n\right]\right)_{n\in\mathbb{N}}$ possesses some ``long-term'' connectivity property, as formalized next.\smallskip\\
\noindent \textbf{Assumption B (On the graph connectivity).}  
The graph sequence $\{\mathcal{G}[n]\}_{n\in\mathbb{N}}$ is $B$-strongly connected, i.e., there exists an integer $B > 0$  (possibly unknown to the agents) such that the graph with edge set $\cup_{t=kB}^{(k+1)B-1} \mathcal{E}[t]$ is strongly connected,  for all $k\geq0$.

\smallskip

\noindent In words, Assumption B says that the information sent by any agent $i$ at any time $n$ will reach any agent $j$ within the next $B$ time slots.

Our goal is to develop an algorithm that converges to
stationary solutions of Problem (\ref{eq: P}) while being implementable
in the above distributed setting (Assumptions A and B), and applicable to arbitrary network topologies without requiring any knowledge of the graph sequence $\mathcal G[n]$. 
%\begin{definition}
%	A point $\mathbf{x}^\star\in\mathcal{K}$ is a stationary solution  of Problem \eqref{eq: P} if there exists a subgradient $G\left(\mathbf{x}^\star\right)$ such that 
%	\begin{equation}
	%	\left(\nabla F\left(\mathbf{x}^\star\right)+\partial G\left(\mathbf{x}^\star \right)\right)^T\left(\mathbf{y}-\mathbf{x}^\star\right)\geq 0,\,\forall \mathbf{y}\in \mathcal{K}.\nonumber
%	\end{equation}
%\end{definition}
To shed light on the core idea of the novel framework, we first introduce  an informal
and constructive description of the proposed algorithm, see Sec. \ref{informal_description}. Sec. \ref{sec: SONATA} will formally introduce SONATA along with its convergence properties.

\subsection{SONATA at a glance}\label{informal_description}
Designing distributed algorithms for Problem (\ref{eq: P}) faces
two main challenges, namely: the nonconvexity of the objective function  and the
lack of global information on the optimization problem from the agents.
To cope with these issues,
SONATA  combines SCA techniques (Step 1 below) with
a consensus-like step implementing a novel broadcast protocol  (Step 2), as described next. 

\noindent\textbf{Step 1: Local SCA.}  Each agent $i$ maintains  a local copy of the common optimization variable $\mathbf{x}$, denoted by $\mathbf{x}_i$, which needs to be updated at each iteration; let $\mathbf{x}_i[n]$ be the value of $\mathbf{x}_i$ at iteration $n$. %Our algorithm has two objectives, namely, all $\mathbf{x}_{i}$'s should be equal asymptotically; and the common value should be a stationary point of Problem \eqref{eq: P}. To this end, SONATA is composed of two major steps: (1) local optimization, and (2) consensus averaging. In the optimization step, each agent updates $\mathbf{x}_{i}$ by moving along its locally computed descent direction; and in the consensus step, agents exchange information  so that the $\mathbf{x}_{i}$'s are forced to agree on a common value.
%Consider solving Problem \eqref{eq: P} from the viewpoint of agent $i$. 
The nonconvexity of $f_i$ together with  the lack of knowledge of  $\sum_{j\neq i} f_j$, prevent  agent $i$ to   solve  Problem \eqref{eq: P} directly.   To cope with this issues, we leverage SCA techniques: at each iteration $n$, agent $i$ solves instead  a convexification of Problem \eqref{eq: P}, having the following form \vspace{-0.1cm}
\begin{equation}
	\widehat{\mathbf{x}}_i\left(\mathbf{x}_i\left[n\right]\right)= \argmin_{\mathbf{x}_i\in \mathcal{K}} \widehat{F}_i\left(\mathbf{x}_i;\mathbf{x}_i\left[n\right]\right) + G\left(\mathbf{x}_i\right), \label{eq: x_hat}\vspace{-0.1cm}
\end{equation}
where the   nonconvex function $F$ is replaced with the  strongly convex approximation $\widehat{F}_i\left(\mathbf{x}_i;\mathbf{x}_i\left[n\right]\right)$ around   $\mathbf{x}_i[n]$, defined as\vspace{-0.1cm}\begin{equation}
	\widehat{F}_i\left(\mathbf{x}_i;\mathbf{x}_i\left[n\right]\right) = \widetilde{f}_i\left(\mathbf{x}_i;\mathbf{x}_i\left[n\right]\right) + \boldsymbol{\pi}_i\left[n\right] ^T\left(\mathbf{x}_i-\mathbf{x}_i\left[n\right]\right),\label{eq: surrogate F_hat}\vspace{-0.1cm}
\end{equation}
 wherein $\widetilde{f}_i\left(\cdot;\mathbf{x}_{i}\left[n\right]\right):\mathcal{K}\to \mathbb{R}$ is a  strongly convex surrogate of the (possibly) nonconvex  $f_i$,  and $\boldsymbol{\pi}_i\left[n\right]$ is the linearization of the unknown term $\sum_{j\neq i} f_j$ around   $\mathbf{x}_i[n]$, i.e.,\vspace{-0.1cm} \begin{equation}
	\boldsymbol{\pi}_i\left[n\right] \triangleq  \sum_{j\neq i}\nabla f_j\left(\mathbf{x}_i\left[n\right]\right).\label{eq: pi}\vspace{-0.2cm}
\end{equation}
 %the  address these two issues, we leverage the idea of SCA: at each iteration, agent $i$ minimizes  a  convex surrogate of $U$ obtained by replacing the (possibly) nonconvex  $f_i$  with a strongly convex approximation $\widetilde{f}_i\left(\cdot;\mathbf{x}_{i}\left[n\right]\right):\mathcal{K}\to \mathbb{R}$, and linearizing the unknown term $\sum_{j\neq i} f_j$

%function of it. In particular, at iteration $n$, we linearize unknown part  $\sum_{j\neq i} f_j$; and  approximate the nonconvex part $f_i$  by a strongly convex function $\widetilde{f}_i\left(\cdot;\mathbf{x}_{i}\left[n\right]\right):\mathcal{K}\to \mathbb{R}$.

%To be precise, each agent $i$ finds a surrogate function of $F$ takes the form
%where
%\begin{equation}
%	\boldsymbol{\pi}_i\left[n\right] = \sum_{j\neq i}\nabla f_j\left(\mathbf{x}_i\left[n\right]\right),\label{eq: pi}
%\end{equation}
%and solves the following \textit{strongly convex} problem:
%\begin{equation}
%	\widehat{\mathbf{x}}_i\left(\mathbf{x}_i\left[n\right]\right)= \arg\min_{\mathbf{x}_i\in \mathcal{K}} \widehat{F}_i\left(\mathbf{x}_i;\mathbf{x}_i\left[n\right]\right) + G\left(\mathbf{x}_i\right). \label{eq: x_hat}
%\end{equation}
Note that  $\widehat{\mathbf{x}}_i\left(\mathbf{x}_i\left[n\right]\right)$ is well-defined, because \eqref{eq: x_hat} has a unique solution.  The direct use of  $\widehat{\mathbf{x}}_i$
 as the new local estimate ${\mathbf{x}}_i[n+1]$ may affect convergence  because 
it might be a too ``aggressive'' update. %; and (ii) we have not
%introduce any mechanism yet to ensure that the local estimates ${\mathbf{x}}_i$
 %eventually agree among all agents. 
 To cope with this issue
 we introduce a step-size in the update of ${\mathbf{x}}_i$:
\begin{equation}\label{wi}
\mathbf{v}_i[n] =\mathbf{x}_i[n] + \alpha[n]\left(\widehat{\mathbf{x}}_i(\mathbf{x}_i[n])- \mathbf{x}_{i}[n]\right), 
\end{equation}
where $\alpha[n]$ is a step-size   (to be properly chosen, see Th. 1). The idea behind the iterates \eqref{eq: x_hat}-\eqref{wi} is to compute stationary
solutions of Problem (\ref{eq: P}) as fixed-points of the mappings
  $\widehat{\mathbf{x}}_i(\bullet)$. To this end, we require the following assumptions on the surrogate function  $\widetilde{f}_i$.

{\noindent\textbf{Assumption C (On the surrogate function).} Each function $\widetilde{f}_i$ satisfies the following properties:
	 \begin{enumerate*}
	 	\item[\textbf{(C1)}] $\nabla \widetilde{f}_i\left(\mathbf{x};\mathbf{x}\right)=\nabla f_i\left(\mathbf{x}\right)$, for all $\mathbf{x}\in \mathcal{K}$;
	 	\item[\textbf{(C2)}] $\widetilde{f}_i\left(\bullet;\mathbf{y}\right)$ is uniformly strongly convex  on $\mathcal{K}$;
	 	\item[\textbf{(C3)}] $\nabla \widetilde{f}_{i}\left(\mathbf{x};\bullet\right)$ is uniformly Lipschitz continuous on $\mathcal{K}$;
	 \end{enumerate*}
}
Conditions C1-C3 are quite natural:  $\widetilde{f}_i$ should be regarded
as a (simple) convex, local, approximation of $f_i$ at the point $\mathbf{x}$
that preserves the first order properties of $f_i$. Several feasible
choices are possible for a given $f_i$;  we
discuss alternative options   in Sec. \ref{sec: discussion}. Here, we only
remark that no extra conditions on  $\widetilde{f}_i$  are required to guarantee
convergence of the proposed algorithm.

 The next proposition 
 establishes the desired connection between the fixed points of  $\widehat{\mathbf{x}}_i(\bullet)$ and the stationary solutions of  \eqref{eq: P}; the proof follows from    \cite[Prop. 8(b)]{facchinei2015parallel} and thus is omitted.  
\begin{proposition}\label{prop: surrogate}
Consider Problem \eqref{eq: P} under Assumptions A1-A6. If the surrogate functions $\widetilde{f}_{i}$'s are chosen according to  Assumption C, then the set of fixed-points of $\widehat{\mathbf{x}}_i(\bullet)$ coincides with that of  stationary solutions of Problem \eqref{eq: P}.
\end{proposition}

\noindent \textbf{Step 2: Broadcasting local information.}  We have now
to introduce a mechanism   to ensure that  the  local estimates ${\mathbf{x}}_i$ eventually agree among all agents. 
  %To force the asymptotic agreement among
%agents' local variables  $\mathbf{x}_i$'s, To do so, a  consensus-like step is employed  on $\mathbf{w}_i$'s. 
To   disseminate information over a  time-varying digraph  without requiring the knowledge of the sequence of digraphs and a double-stochastic weight matrix,   we propose  the following   broadcasting protocol. %, inspired by the push-sum idea \cite{tsianos2012push,tsianos2011distributed,nedic2015distributed,tatarenko2015non}. 
%Note that although push-sum has been used in previous works \cite{tsianos2012push,tsianos2012consensus,nedic2009distributed,zeng2015extrapush},  the algorithms are gradient-based and limited to solving unconstrained problems. In contrast, our algorithm is capable of solving constrained  problems by letting each agent having a different sequence of step-size. In addition, the choice of step-size requires no additional coordination among the agents.
Given $\mathbf{v}_i[n]$, each agent $i$ updates its own local estimate  $\mathbf{x}_i$ together with one extra scalar variable  $\phi_i\left[n\right]$  (initialized to  $\phi_i\left[0\right]=1$), according to
\begin{align}
	\phi_i\left[n+1\right]&=\sum_{j\in\mathcal{N}_i^{\textrm{in}}\left[n\right]}a_{ij}[n]\phi_j\left[n\right];\label{eq: mixing phi}\\
	\mathbf{x}_i\left[n+1\right]&=\dfrac{1}{\phi_i\left[n+1\right]}\sum_{j\in\mathcal{N}_i^{\textrm{in}}\left[n\right]}a_{ij}[n]\phi_j\left[n\right]\mathbf{v}_{j}[n],
	\label{eq: update x}
\end{align}
where the $a_{ij}[n]$'s are some weighting coefficients (to be properly chosen) matching the graph $\mathcal{G}[n]$ in the following sense.%  and satisfy the assumptions as follows.

{\noindent\textbf{Assumption D (On the weighting matrix).} Matrix $\mathbf{A}[n]\triangleq (a_{ij}[n])_{i,j}$ satisfies the following conditions:
		 \begin{enumerate*}
		 	\item[\textbf{(D1)}]  $a_{ii}[n] \geq \kappa>0$ for all $i=1,\ldots,I$ and $n\in\mathbb{N}$;
	\item[\textbf{(D2)}] $a_{ij}[n]\geq \kappa>0$ if $\left(j,i\right)\in \mathcal{E}[n]$, and $a_{ij}=0$ otherwise;
	\item[\textbf{(D3)}] $\mathbf{A}[n]$ is column stochastic, i.e., $\mathbf{1} ^T\mathbf{A}[n]= \mathbf{1}^T$.
	\end{enumerate*}
}

Steps \eqref{eq: mixing phi}-\eqref{eq: update x} are interpreted as follows: All agents   i) send their local variables $\phi_{j}[n]$ and $\phi_{j}[n]\mathbf{v}_{j}[n]$ to their out-neighbors; and ii) linearly combine with coefficients $a_{ij}[n]$ the information coming from their in-neighbors. The idea behind the  use  of the extra variable $\phi_i\left[n\right]$ is to dynamically construct a row stochastic weight matrix  so that consensus among the  $\mathbf{x}_{i}$'s can be asymptotically achieved; see  Sec. \ref{sec: discussion} for more details.

\noindent \textbf{On the local update of $\boldsymbol{\pi}_i\left[n\right]$.} The algorithm developed so far is based on the computation of  $\widehat{\mathbf{x}}_i\left(\mathbf{x}_i\left[n\right]\right)$ in  \eqref{eq: x_hat}. To do so, at each iteration, every agent  $i$  needs to evaluate   $\boldsymbol{\boldsymbol{\pi}}_i\left[n\right]$ and thus  know \emph{locally} all $\nabla f_j(\mathbf{x}_i\left[n\right])$, which is    not feasible in a distributed time-varying setting. %Note that each $\nabla f_j$ in   $\boldsymbol{\boldsymbol{\pi}}_i\left[n\right]$   need to be evaluated in the local variable $\mathbf{x}_i\left[n\right]$, an information that is not available by agent $j$.  We postpone the solution of this issue to the end of this section.
  To cope with this issue, we replace  $\boldsymbol{\pi}_i\left[n\right]$ in \eqref{eq: x_hat} with  an estimate $\widetilde{\boldsymbol{\pi}}_i\left[n\right]$ and solve instead \begin{equation}
	\widetilde{\mathbf{x}}_i\left[n\right] = \argmin_{\mathbf{x}_i\in\mathcal{K}} \widetilde{f}_i\left(\mathbf{x}_i;\mathbf{x}_i\left[n\right]\right) + \widetilde{\boldsymbol{\pi}}_i\left[n\right] ^T\left(\mathbf{x}_i-\mathbf{x}_i\left[n\right]\right) + G\left(\mathbf{x}_i\right).\label{eq: x_tilde}
\end{equation}
  The question now becomes how to update each  $\widetilde{\boldsymbol{\pi}}_i$ using only local
information [in the form of  \eqref{eq: mixing phi}-\eqref{eq: update x}] while asymptotically converging to $\boldsymbol{\pi}_i\left[n\right]$.
%Given that each agent $i$ knows its own gradient information $\nabla f_i\left(\mathbf{x}_i\left[n\right]\right)$, we first
As in \cite{Lorenzo2016NEXT}, rewriting first $\boldsymbol{\pi}_i\left[n\right]$ as
\begin{equation}\label{pi_v2}
	\boldsymbol{\pi}_i\left[n\right] = I\cdot \overline{\nabla f}\left(\mathbf{x}_i\left[n\right]\right) -\nabla f_i\left(\mathbf{x}_i\left[n\right]\right),
\end{equation}
with  $\overline{\nabla f}\left(\mathbf{x}_i\left[n\right]\right) \triangleq  \frac{1}{I}\sum_{j=1}^{I}\nabla f_j\left(\mathbf{x}_i\left[n\right]\right)$, we propose to update   $\widetilde{\boldsymbol{\pi}}_i$ mimicking (\ref{pi_v2}):
 \begin{equation}
	\widetilde{\boldsymbol{\pi}}_i\left[n\right] = I \cdot \mathbf{y}_i\left[n\right] - \nabla f_i\left(\mathbf{x}_i\left[n\right]\right).\label{eq: pi_tilde}
\end{equation}
where $\mathbf{y}_i\left[n\right] $ is  a local variable (controlled by agent $i$)
whose task is to asymptotically track $\overline{\nabla f}\left(\mathbf{x}_i\left[n\right]\right)$. Similar to  \eqref{eq: mixing phi}-\eqref{eq: update x}, we propose the  following new gradient tracking step:
	\begin{align}
	&\mathbf{y}_i \left[n+1\right]= \dfrac{1}{\phi_i\left[n+1\right]}\times\nonumber\\
	& \left(\sum_{j\in\mathcal{N}_i^{\textrm{in}}\left[n\right]}a_{ij}\left[n\right]\phi_j\left[n\right]  \mathbf{y}_j \left[n\right] + \nabla f_i\left(\mathbf{x}_i\left[n+1\right]\right) - \nabla f_i\left(\mathbf{x}_i\left[n\right]\right)\right).\label{eq: update y} 
	\end{align}
where $\phi_i\left[n+1\right]$ is defined in \eqref{eq: mixing phi}. Note that the update of $\mathbf{y}_i$ and thus of $\boldsymbol{\pi}_i[n]$ can be now performed locally by agent $i$, with the same signaling as for \eqref{eq: mixing phi}-\eqref{eq: update x}.

\subsection{Successive Convex Approximation over Time-varying Digraphs}\label{sec: SONATA}
We are now in the position to formally introduce  SONATA, as given   in Algorithm \ref{alg: SONATA}, whose convergence is  stated in Theorem  \ref{thm: convergence}.
\begin{theorem}[\cite{UnboundedGradient}]\label{thm: convergence}
	Let  $\left(\{\mathbf{x}_i\left[n\right]\}_{i=1}^{I}\right)_n$  be the sequence generated by Algorithm 1, and let $\{\bar{\mathbf{z}}[n]\triangleq \left(1/I\right)\sum_i \phi_{i}[n]\cdot\mathbf{x}_i\left[n\right]\}_n$.  Suppose that i) Assumptions A-D hold; ii) the step-size sequences $\{\alpha[n]\}_n$ satisfying  $\alpha\left[n\right]\in \left(0,1\right]$ and  $\sum_{n=0}^{\infty} \alpha\left[n\right]=+\infty$. Then,
	
	 \noindent (1) \emph{[}\texttt{convergence}\emph{]}: $\bar{\mathbf{z}}\left[n\right]$ is bounded for all $n$, and every limit point of $\bar{\mathbf{z}}\left[n\right]$ is a stationary  solution of Problem \eqref{eq: P};\\
	 \noindent (2) \emph{[}\texttt{consensus}\emph{]}:  $\|\mathbf{x}_i\left[n\right]-\bar{\mathbf{z}}\left[n\right]\|\to 0$ as $n\to +\infty$, for all $i$. \label{thm: convergence}
\end{theorem}
\begin{remark}[On the convergence conditions] \emph{ 
We point out  that convergence of SONATA (as stated in Th.\,1)  can also be established under weaker assumptions, namely: i)  a \emph{constant} step-size, (possibly) different for each agent, can be used; and ii) Assumption A4 is not needed. We refer the reader to  \cite{UnboundedGradient} for the proof.}\vspace{-0.2cm}
\end{remark}
\begin{algorithm}[t]
	\SetAlgoLined
	\KwData{For all agent $i$, $\mathbf{x}_i\left[0\right]\in \mathcal{K}$, $\phi_i\left[0\right]=1$, $\mathbf{y}_i\left[0\right]=\nabla f_i\left(\mathbf{x}_i\left[0\right]\right)$, $\widetilde{\boldsymbol{\pi}}_i\left[0\right] = I\mathbf{y}_i\left[0\right]-\nabla f_i\left(\mathbf{x}_i\left[0\right]\right)$. Set $n=0$.}

		 \CommentSty{[S.1]} If $\mathbf{x}_{i}[n]$	satisfies termination criterion: STOP;\\
		 \CommentSty{[S.2] Distributed Local SCA:}		
		 Each agent $i$: \\
		 \Indp (a) computes $\widetilde{\mathbf{x}}_i\left[n\right]$ with \eqref{eq: x_tilde};\\
		 (b) updates its local variable $\mathbf{v}_{i}$ with \eqref{wi} (replace $\widehat{\mathbf{x}}\left(\bullet\right)$ by $\widetilde{\mathbf{x}}\left(\bullet\right)$).
	
	\Indm \CommentSty{[S.3] Consensus: }
		 	 Each agent $i$ broadcasts its local variables and sums up the received variables:\\
		\Indp 	 (a) Update $\phi_i\left[n+1\right]$ with \eqref{eq: mixing phi}.\\
			 (b) Update  $\mathbf{x}_i\left[n+1\right]$ with \eqref{eq: update x}.\\
			 (c) Update   $\mathbf{y}_i\left[n+1\right]$ with  \eqref{eq: update y}.\\
			(d) Update $\widetilde{\boldsymbol{\pi}}_i\left[n+1\right]$ with \eqref{eq: pi_tilde}.\\
		\Indm \CommentSty{[S.4] } $n\longleftarrow n+1$, go to \CommentSty{[S.1] }

	\caption{Successive Convex Approximation over Time-varying Digraphs (SONATA)}\label{alg: SONATA}
\end{algorithm}
%In words,  Theorem \ref{thm: convergence} states that  the weighted average of the $\mathbf{x}_i\left[n\right]$'s, $\bar{\mathbf{z}}\left[n\right]$, will be driven to a stationary solution  of   \eqref{eq: P} while the local agents' variables  $\mathbf{x}_i\left[n\right]$  will asymptotically all agree to  $\bar{\mathbf{z}}\left[n\right]$.\vspace{-0.2cm}% asymptotically. In other words, every agent will eventually obtain a same stationary point of Problem \ref{eq: P}.
 \subsection{Discussion on Algorithm 1}\label{sec: discussion}
   \noindent\textbf{ATC- versus CTA-based updates.}  %We provide next some insight on the algorithm  dynamics that  sheds light on its convergence. %The purpose of the updates ()-() is to force the local variable $\mathbf{x}_{i}[n]$'s to agree on their weighted average $\bar{\mathbf{z}}[n]$, while $\bar{\mathbf{z}}[n]$ is converging to a stationary solution of Problem \eqref{eq: P}. 
 % Theorem \ref{thm: convergence} states that  the weighted average of the $\mathbf{x}_i\left[n\right]$'s, $\bar{\mathbf{z}}\left[n\right]$, will be driven to a stationary solution  of   \eqref{eq: P} while the local agents' variables  $\mathbf{x}_i\left[n\right]$  will asymptotically all agree to  $\bar{\mathbf{z}}\left[n\right]$. We provide next some insight on these two steps. 
 To illustrate the algorithm dynamics, let us combine   \eqref{wi}-\eqref{eq: update x}. Eliminating the auxiliary variable $\mathbf{v}_{i}[n]$, one can write %  one can see that the $\mathbf{x}_{i}\left[n\right]$'s follow the dynamic
 \vspace{-0.2cm}
 \begin{equation}
 \hspace{-0.2cm}{\bf x}_{i}\left[n+1\right]=\sum_{j=1}^{I} w_{ij}\left[n\right]\left({\bf x}_{j}\left[n\right]+\alpha\left[n\right]\left(\widetilde{\mathbf{x}}_j\left(\mathbf{x}_j\left[n\right]\right)-{\bf x}_{j}\left[n\right]\right)\right),\label{eq: x dynamic}\vspace{-0.2cm}
 \end{equation}
 where $\mathbf{W}\left[n\right]\triangleq (w_{ij}[n])_{i,j}$ is a nonnegative matrix with elements
 \begin{equation}\label{eq:def W}
 w_{ij}\left[n\right]=
 \begin{cases}
 \dfrac{a_{ij}\left[n\right]\phi_{j}\left[n\right]}{\sum_{j}a_{ij}\left[n\right]\phi_{j}\left[n\right]}, & \forall j\in\mathcal{N}_{i}^{\textrm{in}}\left[n\right]\\
 0, & \textrm{otherwise.}
 \end{cases} 
 \end{equation}
 Eq. \eqref{eq: x dynamic}  follows an  Adapt-Then-Combine-based (ATC) scheme, where each agent $i$ first updates its local copy $\mathbf{x}_i[n]$ along the ``descent direction'' $\widetilde{\mathbf{x}}_i\left(\mathbf{x}_i\left[n\right]\right)-{\bf x}_{i}\left[n\right]$, and then it combines its new update with that of its neighbors  via consensus, using the weights $\left\{w_{ij}[n]\right\}_{j\in \mathcal N_i[n]}$.
 
 As an alternative to Eq. \eqref{eq: x dynamic}, one can also  follow a so-called Combine-Then-Adapt-based (CTA) approach: each agent $i$ first mixes its own local copy $\mathbf{x}_i[n]$ with that of its neighbors via consensus, and  then it performs its local optimization-based update. The CTA scheme  yields the following alternative:
 \begin{equation}\label{eq: x dynamic-CTA}
 {\bf x}_{i}\left[n+1\right]=\sum_{j=1}^{I} w_{ij}\left[n\right]{\bf x}_{j}\left[n\right]+\alpha\left[n\right]\left(\widetilde{\mathbf{x}}_i\left(\mathbf{x}_i\left[n\right]\right)-{\bf x}_{i}\left[n\right]\right).\vspace{-0.2cm}
 \end{equation}
 We remark that SONATA based on CTA updates is proved to converge under the same conditions as in Theorem\,1 (and Remark\,1); see \cite{UnboundedGradient}.

 \noindent\textbf{On the choice of the surrogate functions.}
SONATA represents a gamut of algorithms, each of them corresponding to a specific choice of the surrogate function $\tilde{f}_i$ and step-size $\alpha[n]$. Some instances of valid $\tilde{f}_i$'s are given next, see \cite{facchinei2015parallel,Lorenzo2016NEXT} for more examples. 

\noindent \textit{$-$Linearization:} When there is no convex structure to exploit, one can simply linearize $f_{i}$, which leads to 
\begin{align}
&\widetilde{f}_{i}\left(\mathbf{x}_{i};\mathbf{x}_{i}\left[n\right]\right) = f_{i}\left(\mathbf{x}_{i}\left[n\right]\right)+  \nabla f_{i}\left(\mathbf{x}_{i}\left[n\right]\right)^{T}\left(\mathbf{x}_{i}-\mathbf{x}_{i}\left[n\right]\right)\nonumber\\
&+\frac{\tau_{i}}{2}\|\mathbf{x}_{i}-\mathbf{x}_{i}\left[n\right]\|^{2}. \label{eq: linearization}
\end{align}
In this case, SONATA becomes a distributed proximal gradient algorithm for constrained optimization.

\noindent \textit{$-$Partial Linearization:} Consider the case that $f_{i}$ can be decomposed as $f_{i}\left(\mathbf{x}_{i}\right)=f_{i}^{\left(1\right)}\left(\mathbf{x}_{i}\right)+f_{i}^{\left(2\right)}\left(\mathbf{x}_{i}\right)$, where $f_{i}^{\left(1\right)}$ is convex and $f_{i}^{\left(2\right)}$ is  nonconvex with Lipschitz continuous gradient.  Preserving the convex part of $f_i$ while linearizing $f_{i}^{\left(2\right)}$  leads to the following valid surrogate  
\begin{align}
\widetilde{f}_{i}\left(\mathbf{x}_{i};\mathbf{x}_{i}\left[n\right]\right) = &f_{i}^{\left(1\right)}\left(\mathbf{x}_{i}\right) + f_{i}^{\left(2\right)}\left(\mathbf{x}_{i}\left[n\right]\right)+ \frac{\tau_{i}}{2}\|\mathbf{x}_{i}-\mathbf{x}_{i}\left[n\right]\|^{2} \nonumber\\
&  + \nabla f_{i}^{\left(2\right)}\left(\mathbf{x}_{i}\left[n\right]\right)^{T}\left(\mathbf{x}_{i}-\mathbf{x}_{i}\left[n\right]\right).\label{eq: partial linearization}
\end{align}
\noindent \textit{$-$Convexification:} If variable $\mathbf{x}_{i}$ can be partitioned as $(\mathbf{x}_{i}^{(1)},\mathbf{x}_{i}^{(2)})$, and $f_{i}$ is convex with respect to $\mathbf{x}_{i}^{(1)}$ while nonconvex with respect to $\mathbf{x}_{i}^{(2)}$,  then $\widetilde{f}_{i}$ can be constructed by convexifying only  the nonconvex part of $f_{i}$, i.e., 
\begin{align}
\widetilde{f}_{i}\left(\mathbf{x}_{i};\mathbf{x}_{i}\left[n\right]\right)=&f_{i}\left(\mathbf{x}_{i}^{(1)},\mathbf{x}_{i}^{(2)}\left[n\right]\right)+\frac{\tau_{i}}{2}\|\mathbf{x}_{i}^{(2)}-\mathbf{x}_{i}^{(2)}\left[n\right]\|^{2}\\
&+\nabla^{(2)}f_{i}\left(\mathbf{x}_{i}[n]\right)^{T}\left(\mathbf{x}_{i}^{(2)}-\mathbf{x}_{i}^{(2)}\left[n\right]\right),
\end{align}
where $\nabla^{(2)}f_{i}\left(\bullet\right)$ is the gradient of $f_{i}$ with respect to $\mathbf{x}_{i}^{(2)}$.
 
 %Other  sophisticated choices of surrogates are omitted due to space limitation. We refer the readers to \cite{Lorenzo2016NEXT} for the details.

  \noindent\textbf{On the choice of the step-size.}    Th. \ref{thm: convergence} offers some
flexibility in the choice of the step-size $\alpha\left[n\right]$ sequence; the conditions therein  ensure that the sequence decays to
zero, but not too fast. There are many diminishing step-size
rules in the literature satisfying the aforementioned  conditions; see, e.g., \cite{bertsekas1999nonlinear}. We found  the following two choices effective in our experiments:  
\begin{equation}\alpha\left[n\right] = \alpha_{0}/\left(n+1\right)^{\beta}, \quad\alpha_0>0,\quad 0.5<\beta\leq 1; \label{alpha_rule_1} \end{equation}
\begin{equation}\alpha\left[n\right] = \alpha\left[n-1\right]\left(1-\mu\alpha\left[n-1\right] \right),\alpha[0]\in\left(0,1\right], \mu\in\left(0,1\right).\label{alpha_rule_2} \end{equation}

{\noindent\textbf{On the choice of matrix $\mathbf{A}[n]$.}   The key requirement of Assumption D is that each  $\mathbf{A}[n]$ is column stochastic. To the best of our knowledge, this is the weakest condition on the weighting matrix to solve optimization problems over arbitrary time-varying digraphs. We remark that our protocol contains push-sum \cite{kempe2003gossip} as a special case if $\mathbf{A}[n]$ is chosen as
	\begin{align}\label{eq: push sum weighting}
		a_{ij}[n] = \begin{cases}
		\dfrac{1}{d_j[n]} & \left(j,i\right)\in \mathcal{E}[n],\\
		0 & \textrm{otherwise}.
		\end{cases}
	\end{align}
Note that the message passing  protocol based on \eqref{eq: push sum weighting} can be easily implemented, since each agent only needs the know its out-degree and broadcast the information evenly to all its out-neighbors. }
	
{Finally, we observe that if the graph is undirected, then $\mathbf{A}[n]$ satisfying Assumption D can chosen to be  double-stochastic. In this case, $\phi_i[n] = 1$ for all $i=1,\ldots,I$ and $n\in\mathbb{N}$; hence step \eqref{eq: mixing phi} in the algorithm can be eliminated. In practice,  rules such as the uniform weights \cite{blondel2005convergence}, Laplacian weights \cite{scherber2004locally}, and Metropolis-Hastings weights \cite{xiao2005scheme} can be adopted to assign $\mathbf{A}[n]$.}\section{SONATA and Special Cases}\label{sec:connection}
In this section we contrast SONATA with related algorithms proposed in the literature \cite{Lorenzo2015NEXT-CAMSAP15,Lorenzo2015NEXT-ICASSP16,Lorenzo2016NEXT,Xu2015augmented}   (including very recent proposals \cite{nedich2016achieving,XiKha-J'16,qu2016harnessing}, appeared online after the submission of this work) for \emph{special}   instances of Problem (1). Specifically, we show next that all these schemes are special cases of SONATA. % describe the connection of SONATA and a few recently proposed distributed algorithms leveraging the technique of gradient tracking. 
To this end, we first rewrite SONATA in an equivalent more convenient form and provide some specific instances of the main algorithm.

\subsection{Preliminaries: SONATA-NEXT and SONATA-L}
Given Algorithm 1, define \vspace{-0.1cm}%the following notation rewrite SONATA in the following equivalent form For analysis convenience, we introduce the following notation:
\begin{align*}
\boldsymbol{\phi}[n] & \triangleq \left[\phi_{1}[n],\ldots,\phi_{I}[n]\right]^{\top}\\
\boldsymbol{\Phi}[n] & \triangleq \textrm{Diag}\left(\boldsymbol{\phi}[n]\right)\\
\widehat{\boldsymbol{\Phi}}[n] & \triangleq \boldsymbol{\Phi}[n]\otimes \mathbf{I}_{m}\\
\widehat{\mathbf{A}}[n] & \triangleq \mathbf{A}[n]\otimes \mathbf{I}_{m}\\
\widehat{\mathbf{W}}[n] & \triangleq \mathbf{W}[n]\otimes \mathbf{I}_{m},
\end{align*}
where $\mathbf{W}[n]$ is  given in \eqref{eq:def W}, and   $\textrm{Diag}\left(\boldsymbol{\phi}[n]\right)$ denotes  a diagonal matrix whose diagonal entries are the components of the vector $\boldsymbol{\phi}[n]$. %being the main diagonal elements. 
Furthermore, let us concatenate  all the local copies $\mathbf{x}_{i}[n]$s in the $m\,I$-length column vector  $\mathbf{x}[n]\triangleq  \left[\mathbf{x}_{1}[n]^T,\ldots,\mathbf{x}_{I}[n]^T\right]^{T}$; the vector $\mathbf{y}[n]$ is similarly defined. % The same convention applies to other local variables.  
Finally, let
$\mathbf{g}_i[n]  \triangleq   \nabla f_{i}\left(\mathbf{x}_{i}[n]\right)$
and $\Delta\mathbf{x}[n]   \triangleq \widetilde{\mathbf{x}}[n]-\mathbf{x}[n]$.
Using  the above notation, the (ATC- and CTA-based) updates of SONATA  [cf.\,Algorithm 1 and Eq. \eqref{eq: x dynamic-CTA}] can be rewritten in compact form as
\begin{subequations}\label{eq:SONATA matrix form}
	\begin{align}
	\boldsymbol{\phi}[n+1] &= \mathbf{A}[n]\boldsymbol{\phi}[n] \label{eq: update vec phi}\\
	\mathbf{W}[n] &=\boldsymbol{\Phi}[n+1]^{-1}\mathbf{A}[n]\boldsymbol{\Phi}[n]\label{eq: update W}\\
	\mathbf{x}[n+1] &=\widehat{\mathbf{W}}[n]\left(\mathbf{x}[n]+\alpha[n]\Delta\mathbf{x}[n]\right)  \text{ (ATC-based update)} \label{eq: update vec x}\\[-2\jot] 
	\intertext{\begin{equation}\left(\text{or } \mathbf{x}[n+1] =  \widehat{\mathbf{W}}[n] \mathbf{x}[n]+\alpha[n]\Delta\mathbf{x}[n] \text{ (CTA-based update)}\right) \end{equation} }\\[-6ex]
	\mathbf{y}[n+1] & =\widehat{\mathbf{W}}[n]\mathbf{y}[n]+\widehat{\boldsymbol{\Phi}}[n+1]^{-1}\left(\mathbf{g}[n+1]-\mathbf{g}[n]\right).\label{eq: update vec y}	
	\end{align}
\end{subequations}

When the digraphs $\mathcal{G}[n]$ admit a \emph{double-stochastic}  matrix $\mathbf{A}[n]$, and $\mathbf{A}[n]$ in (\ref{eq: update vec phi}) is chosen so, %one can choose a \emph{double-stochastic} matrix $\mathbf{A}[n]$ (, (\ref{eq:SONATA matrix form}) 
the iterates (\ref{eq:SONATA matrix form})  can be further simplified. Indeed, it follows from \eqref{eq: update vec phi} and \eqref{eq: update W}   that $\boldsymbol{\phi}[n] = \mathbf{1}$ and $\mathbf{A}[n] = \mathbf{W}[n]$, for all $n$; and then SONATA in \eqref{eq:SONATA matrix form} reduces to
\begin{subequations}\label{eq:NEXT matrix form}
	\begin{align}
	\mathbf{x}[n+1] &=\widehat{\mathbf{W}}[n]\left(\mathbf{x}[n]+\alpha[n]\Delta\mathbf{x}[n]\right) \text{ (ATC-based update)}  \label{ATC-Step}\\[-2ex]
	\intertext{	\begin{equation}
		\left(\text{or } \mathbf{x}[n+1] =  \widehat{\mathbf{W}}[n] \mathbf{x}[n]+\alpha[n]\Delta\mathbf{x}[n] \text{ (CTA-based update)} \right)
		\end{equation} }\\[-4ex]
	\mathbf{y}[n+1] & =\widehat{\mathbf{W}}[n]\mathbf{y}[n]+\mathbf{g}[n+1]-\mathbf{g}[n]\label{NEXT-track}.
	\end{align}

\end{subequations}
The ATC-based updates (\ref{ATC-Step}) and (\ref{NEXT-track})   coincide  with our previous algorithm NEXT, introduced in \cite{Lorenzo2015NEXT-CAMSAP15,Lorenzo2015NEXT-ICASSP16,Lorenzo2016NEXT}. We will refer to (\ref{eq:NEXT matrix form}) as \emph{(ATC/CTA-)SONATA-NEXT}.

%We show now that \noindent \textbf{NEXT\,(\cite{Lorenzo2015NEXT-CAMSAP15,Lorenzo2015NEXT-ICASSP16,Lorenzo2016NEXT}).} Suppose $\mathbf{A}[n]$ is doubly-stochastic, from \eqref{eq: update vec phi} and \eqref{eq: update W} it can be verified that $\boldsymbol{\phi}[n] = \mathbf{1}$ and $\mathbf{A}[n] = \mathbf{W}[n]$ for all $n$. 

We conclude this section, introducing another special instance of  SONATA, tailored to  Problem  (\ref{eq: P}), when $\mathcal{K} = \mathbb{R}^{m}$ (unconstrained) and $G = 0$ (only smooth objectives). Choose each $\widetilde{f}_{i}$ as first order approximation of $f_i$ (plus a quadratic term), that is,
%Next, we  establish the connections between algorithm SONATA and the  following  gradient-based distributed algorithms for  \emph{unconstrained smooth convex} problems. 
%To this end, we first derive the explicit expression of $\widetilde{\mathbf{x}}_i[n]$ as follows. 
%Letting $\mathcal{K} = \mathbb{R}^{m}$ and $G \equiv 0$, and restricting 
%the  choice of $\widetilde{f}_{i}$ to be the \emph{linearization} scheme, i.e., 
\begin{equation}
\begin{aligned}
\widetilde{f}_{i}\left(\mathbf{x}_{i};\mathbf{x}_{i}[n]\right) = & f_{i}\left(\mathbf{x}_{i}[n]\right) + \nabla f_{i}\left(\mathbf{x}_{i}[n]\right)^{\top}\left(\mathbf{x}_{i} - \mathbf{x}_{i}[n]\right) \\
&+ \frac{\tau_{i}}{2}\norm{\mathbf{x}_{i} - \mathbf{x}_{i}[n]}^{2},
\end{aligned}
\end{equation}
and set $\tau_i=I$. Then, $\widetilde{\mathbf{x}}_{i}[n]$  can be computed in closed form   [cf.\,(\ref{eq: x_tilde})]:
\begin{equation}\label{x_tilde_linearized}
\begin{aligned}
\widetilde{\mathbf{x}}_{i}[n] 
&  = \argmin_{\mathbf{x}_{i}}\, \left(I\cdot\mathbf{y}_{i}[n]\right)^{\top}\left(\mathbf{x}_{i} - \mathbf{x}_{i}[n]\right) + \frac{I}{2}\norm{\mathbf{x}_{i} - \mathbf{x}_{i}[n]}^{2}\\
& = \argmin_{\mathbf{x}_{i}}\, \frac{I}{2}\norm{\mathbf{x}_{i} - \mathbf{x}_{i}[n] +  \mathbf{y}_{i}[n]}^{2}\\
& = \mathbf{x}_{i}[n] - \mathbf{y}_{i}[n].
\end{aligned} 
\end{equation}
Using \eqref{x_tilde_linearized} in (\ref{eq:SONATA matrix form}), we get   
\begin{align}\label{eq:SONATA-L}
\boldsymbol{\phi}[n+1] &= \mathbf{A}[n]\,\boldsymbol{\phi}[n] \nonumber\\
\mathbf{W}[n]&=\boldsymbol{\Phi}[n+1]^{-1}\mathbf{A}[n]\,\boldsymbol{\Phi}[n]\nonumber\\
\mathbf{x}[n+1] & = \widehat{\mathbf{W}}[n]\left(\mathbf{x}[n] - \alpha[n]\,\mathbf{y}[n]\right) \text{ (ATC-based update)}\\[-1.5ex]
\intertext{\begin{equation*}
	\hspace{-1em}\left(\text{or }\mathbf{x}[n+1]  = \widehat{\mathbf{W}}[n]\mathbf{x}[n] - \alpha[n]\,\mathbf{y}[n]\text{ (CTA-based update)}\right)
	\end{equation*}}\\[-6.5ex]
\mathbf{y}[n+1] & =\widehat{\mathbf{W}}[n]\,\mathbf{y}[n]+\widehat{\boldsymbol{\Phi}}[n+1]^{-1}\left(\mathbf{g}[n+1]-\mathbf{g}[n]\right),\nonumber
\end{align}
which we will  refer to as \emph{(ATC/CTA-)SONATA-L} (L stands for ``linearized''). 

%We mention that our convergence proof still hold if the x-update of SONATA follows a  combine-then-adapt (CTA) scheme, i.e., 
%\begin{equation}\label{eq:SONATA-CTA}
%\mathbf{x}[n+1]  = \widehat{\mathbf{W}}[n]\mathbf{x}[n] - \alpha[n]\mathbf{y}[n].
%\end{equation}
Similar to (\ref{eq:NEXT matrix form}), if  all $\mathbf{A}[n]$ are  double stochastic martices, then %$\mathbf{A}[n] = \mathbf{W}[n]$ and 
(ATC/CTA-)SONATA-L  reduces to 
\begin{align}\label{eq:NEXT-L}
\mathbf{x}[n+1] & = \widehat{\mathbf{W}}[n]\left(\mathbf{x}[n] - \alpha\mathbf{y}[n]\right)\text{ (ATC-based update)}\\[-1.5ex]
\intertext{\begin{equation*}
	\hspace{-1em}\left(\text{or }\mathbf{x}[n+1]  = \widehat{\mathbf{W}}[n]\mathbf{x}[n] - \alpha\mathbf{y}[n] \text{ (CTA-based update)}\right)
	\end{equation*}}\\[-6ex]
\mathbf{y}[n+1] & = \widehat{\mathbf{W}}[n]\mathbf{y}[n] + \mathbf{g}[n+1]- \mathbf{g}[n],\nonumber
\end{align}
which is referred to as \emph{(ATC/CTA-)SONATA-NEXT-L}. 
\subsection{Connection with current algorithms}
We are now in the position to show that  the algorithms recently  proposed in \cite{Xu2015augmented, qu2016harnessing,nedich2016achieving,XiKha-J'16} are all special cases of SONATA and NEXT, earlier proposed in \cite{Lorenzo2015NEXT-CAMSAP15,Lorenzo2015NEXT-ICASSP16,Lorenzo2016NEXT}.  Since algorithms in \cite{Xu2015augmented,qu2016harnessing,nedich2016achieving,XiKha-J'16} are applicable  only to \emph{unconstrained} ($\mathcal{K}=\mathbb{R}^{m}$), \emph{smooth} ($G=0$) and \emph{convex} (each $f_i$ is convex) multiagent problems, in the following,  we tacitly consider such an instance of Problem (1).
\vspace{2ex}\\
\textbf{Aug-DGM  \cite{Xu2015augmented} and Algorithm in \cite{qu2016harnessing}.}  Introduced in \cite{Xu2015augmented} for \emph{undirected, time-invariant} graphs, the  Aug-DGM algorithm  reads
\begin{equation}\label{eq:augDGM uncoordinated step-size}
\begin{aligned}
\mathbf{x}[n+1] & = \widehat{\mathbf{W}}\left(\mathbf{x}[n] - \textrm{Diag}\left(\boldsymbol{\alpha}\otimes \mathbf{1}_m\right) \mathbf{y}[n]\right)\\
\mathbf{y}[n+1] & = \widehat{\mathbf{W}}\left(\mathbf{y}[n] + \mathbf{g}[n+1] - \mathbf{g}[n]\right)
\end{aligned}
\end{equation}
where  $\widehat{\mathbf{W}} \triangleq \mathbf{W}\otimes \mathbf{I}_m$,  $\mathbf{W}$ is a double stochastic matrix matching the graph (i.e., $w_{ij}>0$ if $(j,i)\in\mathcal{E}$ and $w_{ij} = 0$ otherwise), and   $\boldsymbol{\alpha}$ is the vector of agents' step-sizes $\alpha_i$s. \\\indent A similar algorithm was proposed in parallel in \cite{qu2016harnessing} (in the same networking setting of \cite{Xu2015augmented}), which reads  
\begin{equation}\label{eq:augDGM coordinated step-size}
\begin{aligned}
\mathbf{x}[n+1] & = \widehat{\mathbf{W}}\left(\mathbf{x}[n]- \alpha\mathbf{y}[n]\right)\\
\mathbf{y}[n+1] & = \widehat{\mathbf{W}}\mathbf{y}[n] + \mathbf{g}[n+1]- \mathbf{g}[n].
\end{aligned}
\end{equation}
While   algorithm \eqref{eq:augDGM uncoordinated step-size} is in principle more general than  \eqref{eq:augDGM coordinated step-size}--agents can use different step-sizes $\alpha_i$s--the assumptions in \cite{Xu2015augmented} on $\boldsymbol{\alpha}$ to guarantee convergence are difficult to be enforced in practice, and in particular in a distributed setting.  % must satisfy to guarantee convergence are not practical, because they call for some global knowledge are not practical satisfy depend not only on the  parameters of the network, but also on the global knowledge of $\boldsymbol{\alpha}$ itself, which makes the scheme less appealing in a distributed computational setting.

Aug-DGM  \cite{Xu2015augmented} was  shown to achieve convergence rate  $O\left(1/n\right)$  for smooth convex functions $f_i$s, and   linear convergence $O\left(\gamma^n\right)$ for some $\gamma \in \left(0,1\right)$, if $f_i$'s are  strongly convex.

Clearly  Aug-DGM  \cite{Xu2015augmented} in  \eqref{eq:augDGM uncoordinated step-size} and Algorithm \cite{qu2016harnessing} in \eqref{eq:augDGM coordinated step-size}   are both  special cases of (ATC-)SONATA-NEXT-L [cf. Eq. \eqref{eq:NEXT-L}].% if $\mathbf{W}[n] \equiv \mathbf{W}$ for all $n$, and the algorithm is shown to achieve a $O\left(1/n\right)$ rate for smooth convex functions and a linear rate $O\left(\gamma^n\right)$ for some $\gamma \in \left(0,1\right)$ if the function is in addition strongly convex. 
\vspace{2ex}

\noindent\textbf{(Push-)DIGing \cite{nedich2016achieving}.} Appeared in the technical report \cite{nedich2016achieving} and applicable to  \emph{$B$-strongly connected undirected graphs}, the DIGing Algorithm   reads
\begin{equation}\label{eq:DIGing}
\begin{aligned}
\mathbf{x}[n+1] & = \widehat{\mathbf{W}}[n]\mathbf{x}[n] - \alpha \mathbf{y}[n]\\
\mathbf{y}[n+1] & = \widehat{\mathbf{W}}[n]\mathbf{y}[n] + \mathbf{g}[n+1]- \mathbf{g}[n],
\end{aligned}
\end{equation}
where $\mathbf{W}[n]$ is a double-stochastic matrix matching the graph.
Clearly, DIGing is a special case of (CTA-)SONATA-NEXT-L [cf. \eqref{eq:NEXT-L}], proposed in the earlier works \cite{Lorenzo2015NEXT-CAMSAP15,Lorenzo2015NEXT-ICASSP16,Lorenzo2016NEXT}.

In the same technical report \cite{nedich2016achieving}, the authors proposed push-DIGing,  the extension of DIGing to  $B$-strongly connected digraphs. It turns out that push-DIGing  is a special case of (ATC-)SONATA-L [cf. Eq. \eqref{eq:SONATA-L}], when $a_{ij}[n] = 1/d_j[n]$. Both DIGing and Push-DIGing are shown to have R-linear convergence rate, when agents' objective functions are strongly convex. %As a byproduct, SONATA is also related to  primal-dual algorithms such as EXTRA \cite{shi2015extra},  which is shown in \cite[Sec. 2.2]{nedich2016achieving} and briefly mentioned next for completeness.
\vspace{2ex}\\
%\noindent \textbf{EXTRA.} Proposed in \cite{shi2015extra}, algorithm EXTRA for static undirected graph takes form
%\begin{equation}\label{eq:EXTRA}
%\begin{aligned}
%\mathbf{x}[n+2]  = \left(\mathbf{I} + \mathbf{W}'\right)\mathbf{x}[n+1] - \widetilde{\mathbf{W}}\mathbf{x}[n] - \alpha \left(\mathbf{g}[n+1] - \mathbf{g}[n]\right).\\
%\end{aligned}
%\end{equation}
%Consider fixed $\mathbf{W}[n]$ and eliminating variable $\mathbf{y}$ in Eq. \eqref{eq:DIGing} leads to the following update:
%\begin{equation}
%	\mathbf{x}[n+2] = 2\mathbf{W}\mathbf{x}[n+1]  - \mathbf{W}^2\mathbf{x}[n] - \alpha\left(\mathbf{g}[n+1] - \mathbf{g}[n]\right),
%\end{equation}
%which fits in Eq. \eqref{eq:EXTRA} with the identification $\mathbf{W}' = 2\mathbf{W} - \mathbf{I}$ and $\widetilde{\mathbf{W}} = \mathbf{W}^2$. 
\noindent \textbf{ADD-OPT \cite{XiKha-J'16}.}  Finally, we mention the  ADD-OPT Algorithm,  proposed in \cite{XiKha-J'16} for \emph{strongly connected static digraphs}, which takes the following form:\vspace{-0.2cm}
\begin{equation}\label{eq:ADD-OPT}
\begin{aligned}
\mathbf{z}[n+1] & = \widehat{\mathbf{A}}\mathbf{z}[n] - \alpha \widetilde{\mathbf{y}}[n]\\
\boldsymbol{\phi}[n+1] & = \mathbf{A}\boldsymbol{\phi}[n]\\
\mathbf{x}[n+1] & = \widehat{\boldsymbol{\Phi}}[n+1]^{-1}\mathbf{z}[n+1]\\
\widetilde{\mathbf{y}}[n+1] & = \widehat{\mathbf{A}}\widetilde{\mathbf{y}}[n] + \mathbf{g}[n+1] - \mathbf{g}[n].
\end{aligned}
\end{equation}
Defining $\mathbf{y}[n] = \widehat{\boldsymbol{\Phi}}[n]^{-1}\widetilde{\mathbf{y}}[n]$, it can be verified that algorithm \eqref{eq:ADD-OPT} can be rewritten as
\begin{equation}
\begin{aligned}
\boldsymbol{\phi}[n+1] &= \mathbf{A}\boldsymbol{\phi}[n]\\
\mathbf{W}&=\boldsymbol{\Phi}[n+1]^{-1}\mathbf{A}\boldsymbol{\Phi}[n]\\
\mathbf{x}[n+1] & = \widehat{\mathbf{W}} \mathbf{x}[n] - \alpha \widehat{\boldsymbol{\Phi}}[n+1]^{-1}\widehat{\boldsymbol{\Phi}}[n] \mathbf{y}[n]\\
\mathbf{y}[n+1] & = \widehat{\mathbf{W}}\mathbf{y}[n] + \widehat{\boldsymbol{\Phi}}[n+1]^{-1}\left(\mathbf{g}[n+1] - \mathbf{g}[n]\right).
\end{aligned}\label{eq:ADD-OPT2}
\end{equation}
Comparing Eq. \eqref{eq:SONATA-L} and \eqref{eq:ADD-OPT2}, one can see that %except for using a fixed mixing matrix $\mathbf{A}$, 
ADD-OPT is an instance of (CTA-)SONATA-L %by switching the order of mixing and descent in the x-update, and uses a 
with the following particular choice of (uncoordinated) step-size: $\frac{\phi_i[n]\alpha}{\phi_i[n+1]}$ for agent $i$. We recall that (CTA-)SONATA-L is guaranteed to converge also with uncordinate step-sizes; see Remark 1.  ADD-OPT is shown to have linear rate $O\left(\gamma^n\right)$ for strongly convex objective functions.

We summarize the connections between the different versions of SONATA(-NEXT) and its special cases  in Table \ref{tab:connection}.
\begin{table*}[]\vspace{-0.2cm}
	\centering
	\caption{Connection of SONATA with current algorithms}
	\label{tab:connection}
	\begin{tabular}{|c|c|c|l|}
		\hline
		\multicolumn{1}{|c|}{Algorithms}            & \multicolumn{1}{c|}{Special cases of}                                                                                     & Instance of Problem (1)                                                                   & \multicolumn{1}{|c|}{Graph topology}                                                                        \\ \hline
		NEXT  \cite{Lorenzo2016NEXT}                            & SONATA \eqref{eq:SONATA matrix form}                                                                                                       & \begin{tabular}[c]{@{}l@{}} $F$ nonconvex\\ $G \neq 0$\\ $\mathcal{K}\subseteq \mathbb{R}^m$\end{tabular} & \begin{tabular}[c]{@{}l@{}}time-varying\\ doubly-stochasticable digraph\end{tabular} \\ \hline
		Aug-DGM         \cite{qu2016harnessing,Xu2015augmented}                  & \begin{tabular}[c]{@{}l@{}}ATC-SONATA-NEXT-L ($\boldsymbol{\alpha}=\alpha\mathbf{1}_I$) \eqref{eq:NEXT-L}\end{tabular}               & \begin{tabular}[c]{@{}l@{}} $F$ convex\\ $G = 0$\\ $\mathcal{K} = \mathbb{R}^m$\end{tabular}     & static undirected graph                                                                \\ \hline
		DIGing \cite{nedich2016achieving}                           & CTA-SONATA-NEXT-L   \eqref{eq:NEXT-L}                                                                                            & \begin{tabular}[c]{@{}l@{}} $F$ convex\\ $G = 0$\\ $\mathcal{K}=\mathbb{R}^m$\end{tabular}     & \begin{tabular}[c]{@{}l@{}}time-varying\\ doubly-stochasticable digraph\end{tabular} \\ \hline
		\multicolumn{1}{|l|}{push-DIGing \cite{nedich2016achieving}} & ATC-SONATA-L   \eqref{eq:SONATA-L}                                                                                       & \begin{tabular}[c]{@{}l@{}} $F$ convex\\ $G = 0$\\ $\mathcal{K}=\mathbb{R}^m$\end{tabular}      & \begin{tabular}[c]{@{}l@{}}time-varying\\ digraph\end{tabular}                         \\ \hline
		ADD-OPT  \cite{XiKha-J'16}                         & \begin{tabular}[c]{@{}l@{}}ATC-SONATA-L \eqref{eq:SONATA-L}\end{tabular} & \begin{tabular}[c]{@{}l@{}} $F$ convex\\ $G = 0$\\ $\mathcal{K}=\mathbb{R}^m$\end{tabular}     & static digraph                                                                         \\ \hline
	\end{tabular}
\end{table*}

\section{Applications and Numerical Results}\label{sec: Numerical}
In this section, we test the performance of SONATA on both convex and nonconvex problems. For all applications, we simulate the following graph topology: at each iteration, each agent has two out-neighbors, with one belonging to a time-varying cycle and the other two randomly chosen. {The step-size $\alpha\left[n\right]$ is chosen based on the rule (\ref{alpha_rule_1}), and matrix $\mathbf{A}[n]$ is chosen based on Eq. \eqref{eq: push sum weighting}.}\vspace{-0.1cm}
\subsection{Robust Regression}
In the first simulation, we consider a robust linear regression problem. Each agent $i$ has $n_i$ measurements of parameter $\mathbf{x}$ as $b_{ij} = \mathbf{a}_{ij}^T\mathbf{x}$, which is corrupted by noise and outliers. To estimate $\mathbf{x}$, we solve the following problem
\begin{equation}
	\begin{aligned}
	&\underset{\mathbf{x}}{\textrm{minimize}}& &\sum_{i=1}^{I}\sum_{j=1}^{n_i}h\left(\mathbf{a}_{ij}^T\mathbf{x}-b_{ij}\right),\\
	\end{aligned}\label{eq: huber loss}
\end{equation}
where $h$ is the Huber loss function given by
\[
h\left(r\right)=\begin{cases}
r^{2}, & \textrm{if }\left|r\right|>c\\
c\left(2\left|r\right|-c\right), & \textrm{if }\left|r\right|\leq c.
\end{cases}
\]
The function behaves like the $\ell_1$-norm if residual $r$ is larger than the cut-off parameter $c$; and is quadratic if $\left|r\right|\leq c$.

Defining $f_i\left(\mathbf{x}\right)\triangleq \sum_{j=1}^{n_{i}}h\left(\mathbf{a}_{ij}^T\mathbf{x}-b_{ij}\right)$,  Problem \eqref{eq: huber loss} is an instance of the general Problem \eqref{eq: P} with $F=\sum_{i=1}^{I}f_i$. We provide two versions of  surrogate function $\widetilde{f}_i$. In the first version, function $f_i$ is linearized at each iteration (cf. Eq. \eqref{eq: linearization}). In the second version, we propose a SCA scheme that approximates $f_i$ at $\mathbf{x}\left[n\right]$ by a quadratic function $\widetilde{f}_i\left(\mathbf{x};\mathbf{x}\left[n\right]\right)=\sum_{j=1}^{n_i}\widetilde{h}_{ij}\left(\mathbf{x};\mathbf{x}\left[n\right]\right)+ \frac{\tau}{2}\|\mathbf{x}-\mathbf{x}\left[n\right]\|^2$, where $\widetilde{h}_{ij}$ is defined as
\[
\widetilde{h}_{ij}\left(\mathbf{x};\mathbf{x}\left[n\right]\right)=\begin{cases}
\frac{c}{r_{ij}\left[n\right]}\left(\mathbf{a}_{ij}^{T}\mathbf{x}-b_{ij}\right) ^2& \textrm{if }\left|r_{ij}\left[n\right]\right|>c\\
c\left(\mathbf{a}_{ij}^{T}\mathbf{x}-b_{ij}\right) ^2& \textrm{if }\left|r_{ij}\left[n\right]\right|\leq c,
\end{cases}
\]
with $r_{ij}\left[n\right]=\mathbf{a}_{ij}^{T}\mathbf{x}\left[n\right]-b_{ij}$. Consequently, the update $\widetilde{\mathbf{x}}_i\left[n\right]$ has a closed form solution given as
$\widetilde{\mathbf{x}}_i\left[n\right]=\left(2\mathbf{A}_i^T\mathbf{D}_i\mathbf{A}_i+\tau\mathbf{I}\right)^{-1}\left(\tau\mathbf{x}\left[n\right]-\widetilde{\boldsymbol{\pi}}_i\left[n\right] + 2\mathbf{A}_i^T\mathbf{D}_i\mathbf{b}_i\right),$
where the $j$th row of $\mathbf{A}_i$ is $\mathbf{a}_{ij}^T$, and the $j$th element of $\mathbf{b}_i$ is $b_{ij}$. Matrix $\mathbf{D}_i$ is diagonal with its $j$th diagonal being $\min\{c,c/r_{ij}\left[n\right]\}$.

We simulate $I=30$ agents collaboratively estimate $\mathbf{x}_0$ of dimension 200 with \textit{i.i.d.} uniformly distributed entries in $\left[-1,1\right]$. Each agent only has  $n_i = 20$ measures. The elements of vector $\mathbf{a}_{ij}$ is generated following an $\textit{i.i.d.}$ Gaussian distribution, then normalized to be $\|\mathbf{a}_{ij}\|=1$. The measurements noise follows a Gaussian distribution with standard deviation $\sigma=0.1$, and  each agent has one measurement corrupted by an outlier following a Gaussian distribution with standard deviation $5\sigma$. The cut-off parameter $c$ is set to be $c=3\sigma$.

Algorithm parameters are tuned as follows. The proximal parameter $\tau$ for our linearization scheme and SCA scheme are set to be $\tau_L=2$ and $\tau_{SCA}=1.5$, respectively. Step-size parameters are set to be $\alpha\left[0\right]=0.1$ and $\mu = 0.01$ for both of them.  We compare the performance with subgradient-push algorithm proposed in \cite{nedic2015distributed}, for which the step-size parameter is set to be $\alpha\left[0\right]=0.5$, $\mu=0.01$. In addition, since SONATA has two consensus steps, we run subgradient-push twice in one iteration using the same graph for a fair comparison.

\begin{figure}
	\centering\vspace{-0.2cm}
	\includegraphics[scale=0.4]{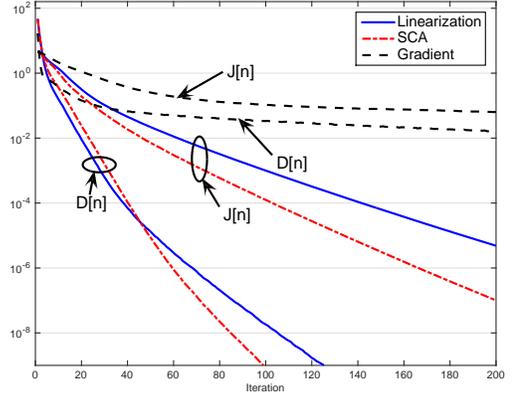}\vspace{-0.2cm}
	\caption{Optimality  measurements  $J\left[n\right]$ and  consensus error $D\left[n\right]$ versus the number of iterations.}
	\label{fig: opt criteria-huber}\vspace{-0.3cm}
\end{figure}

The performance is averaged over 100 Monte-Carlo simulations, where each time $\mathbf{x}_0$ is fixed while the noise and graph connectivity are randomly generated. Fig. \ref{fig: opt criteria-huber}  reports the progress of the algorithms towards optimality and consensus error, where measure $J\left[n\right]$ is defined as $J\left[n\right] \triangleq \|\nabla F\left(\bar{\mathbf{z}}\left[n\right]\right)\|_{\infty}$ and  $D\left[n\right] \triangleq \frac{1}{I} \sum_{i=1}^{I}\|\mathbf{x}_i\left[n\right]-\bar{\mathbf{z}}\left[n\right]\|^2$. We can see that SONATA reaches consensus and convergence much faster than subgradient-push. In addition, SCA scheme outperforms plain linearization by exploiting the convexity of the objective function.

%
%In the second one, we compare with a gradient algorithm proposed in \cite{tatarenko2015non} for distributed target localization problem.
\subsection{Target Localization}
Target Localization problem considers a number of $I$ sensors in a network collaboratively locate the position of $T$ targets. Sensor $i$ has the knowledge of the coordinate of its own location $\mathbf{s}_i$, and the relative Euclidean distance between itself and target $t$, denoted $d_{it}$. The problem is formulated as:
\begin{equation}
	\begin{aligned}
	&\underset{\{\mathbf{x}_t\}_{t=1}^{T}}{\textrm{minimize}}& & \sum_{i=1}^{I}\sum_{t=1}^{T}p_{it}\left(d_{it}-\|\mathbf{x}_t-\mathbf{s}_i\|^2\right)^2\\
	&\textrm{subject to}& &\mathbf{x}_i\in \mathcal{K}\subset\mathbb{R}^m,\,\forall i,\\
	\end{aligned} \label{P: target localization}
\end{equation}
where $\mathcal{K}$ is a compact set and variable $\mathbf{x}_t$ is an estimate of the location of target $t$, denoted $\mathbf{x}_t^0$. Parameter $p_{it}\in\{0,1\}$ takes value zero if the $i$th agent has no measurement about target $t$.

We apply SONATA to Problem \eqref{P: target localization} with $ f_i\left(\mathbf{x}\right)=\sum_{t=1}^{T}p_{it}\left(d_{it}-\|\mathbf{x}_t-\mathbf{s}_i\|^2\right)^2$, where $\mathbf{x}$ is obtained by stacking the $\mathbf{x}_t$'s. The two SCA  schemes proposed in \cite{Lorenzo2016NEXT} are adopted, namely, linearization (cf. Eq. \eqref{eq: linearization}) and partial linearization with surrogate function \vspace{-0.2cm}
\begin{equation}
\widetilde{f}_{i}\left(\mathbf{x};\mathbf{x}\left[n\right]\right) = \sum_{t=1}^{T} p_{it}\left(\widetilde{f}_{it}\left(\mathbf{x};\mathbf{x}\left[n\right]\right)+\frac{\tau}{2}\|\mathbf{x}_{t}-\mathbf{x}_{t}\left[n\right]\|^{2}\right),
\end{equation}
where $\widetilde{f}_{it}\left(\mathbf{x};\mathbf{x}\left[n\right]\right)=\mathbf{x}_{t}^{T}\mathbf{A}_{i}\mathbf{x}_{t}-\mathbf{b}_{it}\left[n\right]^{T}\left(\mathbf{x}_{t}-\mathbf{x}_{t}\left[n\right]\right)$, with $\mathbf{A}_{i}=4\mathbf{s}_{i}\mathbf{s}_{i}^{T}+2\|\mathbf{s}_{i}\|^{2}\mathbf{I}$, and $\mathbf{b}_{it}\left[n\right]=4\|\mathbf{s}_{i}\|^{2}\mathbf{s}_{i}-4\left(\|\mathbf{x}_{t}\left[n\right]\|^{2}-d_{it}\right)\left(\mathbf{x}_{t}\left[n\right]-\mathbf{s}_{i}\right)+8\left(\mathbf{s}_{i}^{T}\mathbf{x}_{t}\left[n\right]\right)\mathbf{x}_{t}\left[n\right]$.

%and consider two successive approximation schemes proposed in \cite{Lorenzo2016NEXT}, namely, plain gradient descent and partial linearization. In the gradient scheme, we obtain a surrogate function by $\widetilde{f}_i\left(\mathbf{x};\mathbf{x}\left[n\right]\right)$ linearizing $f_i$ at point $\mathbf{x}\left[n\right]$; while in the partial linearization scheme, we expand perfect squares and linearize the third and fourth order terms. We omit the details due to space limitation, and refer the readers to \cite{Lorenzo2016NEXT} for the expressions of $\widetilde{f}_i$.

In the simulation, we set the number of sensors to be $I=30$, and the number of targets to be $t=5$. Parameter $p_{it}$ takes value zero and one with equal probability. The locations of the sensors and targets are uniformly randomly generated in $\left[0,1\right]^2$.   We consider a noisy environment that the measured distances are corrupted by $\textit{i.i.d.}$ Gaussian noise. The noise standard deviation is set to be  the minimum pairwise distance between sensors and targets. 

We compare with the gradient algorithm proposed in \cite{tatarenko2015non} for unconstrained optimization. Algorithm parameters are tuned as follows. For our algorithm, the step-size parameters are set to be  $\alpha\left[0\right]=0.1$ and $\mu = 10^{-4}$. The proximal parameter $\tau$ of $\widetilde{f}_i$ for the linearization scheme is selected to be $\tau_L=7$ and that for partial linearization is selected to be $\tau_{PL}=5$. For the benchmark algorithm, $\alpha\left[0\right] = 0.05$ and $\mu=10^{-4}$. 

\begin{figure}\vspace{-0.5cm}
	\centering
	\includegraphics[scale=0.4]{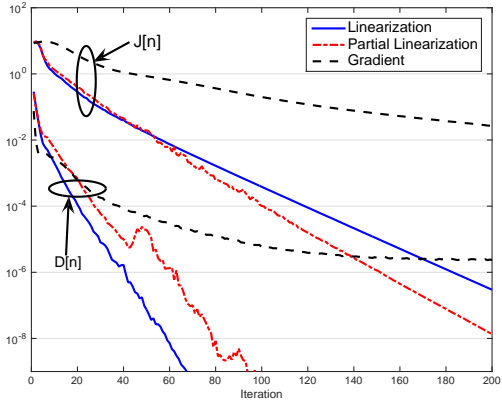}\vspace{-0.2cm}
	\caption{Optimality measurement $J\left[n\right]$ and consensus error $D\left[n\right]$ versus the number of iterations.}\vspace{-0.3cm}
	\label{fig: opt criteria}
\end{figure}

A comparison of the algorithms is given in Fig. \ref{fig: opt criteria}, which is averaged over 100 Monte-Carlo simulations. %To be specific, the position of the sensors and targets are fixed, while the noise and graph connectivity are randomly generated in each trial. 
Fig. \ref{fig: opt criteria}  shows that within 200 iteration, both consensus and convergence are achieved for all algorithms; and SONATA converges much faster than the benchmark gradient algorithm. \vspace{-0.2cm}

\section{Conclusion}\label{sec: conclusion}
In this paper we have proposed (ATC/CTA-)SONATA, a family of novel  distributed algorithms for \emph{nonconvex constrained} optimization over time-varying (directed) networks. The algorithm leverages the idea of SCA for local optimization, a tracking mechanism to locally estimate the gradients of agents' functions, and a new in-network broadcast protocol to distribute the computation and sharing information among agents. SONATA is the first broadcast-based algorithm framework that  can solve  convex or nonconvex constrained optimization problems over arbitrary time-varying digraphs.  SONATA was also shown to contain, as special cases, current algorithms proposed in simplified settings.    Numerical result shows that our algorithm outperforms   state-of-the-art schemes on  considered convex and nonconvex  applications.  \vspace{-0.2cm}

\footnotesize
\bibliographystyle{IEEEtran}
\bibliography{sonata_refs.bib}

\begin{thebibliography}{10}
\providecommand{\url}[1]{#1}
\csname url@samestyle\endcsname
\providecommand{\newblock}{\relax}
\providecommand{\bibinfo}[2]{#2}
\providecommand{\BIBentrySTDinterwordspacing}{\spaceskip=0pt\relax}
\providecommand{\BIBentryALTinterwordstretchfactor}{4}
\providecommand{\BIBentryALTinterwordspacing}{\spaceskip=\fontdimen2\font plus
\BIBentryALTinterwordstretchfactor\fontdimen3\font minus
  \fontdimen4\font\relax}
\providecommand{\BIBforeignlanguage}[2]{{%
\expandafter\ifx\csname l@#1\endcsname\relax
\typeout{** WARNING: IEEEtran.bst: No hyphenation pattern has been}%
\typeout{** loaded for the language `#1'. Using the pattern for}%
\typeout{** the default language instead.}%
\else
\language=\csname l@#1\endcsname
\fi
#2}}
\providecommand{\BIBdecl}{\relax}
\BIBdecl

\bibitem{nedic2009distributed}
A.~Nedi{\'c} and A.~Ozdaglar, ``Distributed subgradient methods for multi-agent
  optimization,'' \emph{IEEE Transactions on Automatic Control}, vol.~54,
  no.~1, pp. 48--61, 2009.

\bibitem{shi2015extra}
W.~Shi, Q.~Ling, G.~Wu, and W.~Yin, ``Extra: An exact first-order algorithm for
  decentralized consensus optimization,'' \emph{SIAM Journal on Optimization},
  vol.~25, no.~2, pp. 944--966, 2015.

\bibitem{srivastava2011distributed}
K.~Srivastava and A.~Nedi{\'c}, ``Distributed asynchronous constrained
  stochastic optimization,'' \emph{IEEE Journal of Selected Topics in Signal
  Processing}, vol.~5, no.~4, pp. 772--790, 2011.

\bibitem{duchi2012dual}
J.~C. Duchi, A.~Agarwal, and M.~J. Wainwright, ``Dual averaging for distributed
  optimization: convergence analysis and network scaling,'' \emph{IEEE
  Transactions on Automatic control}, vol.~57, no.~3, pp. 592--606, 2012.

\bibitem{gharesifard2014distributed}
B.~Gharesifard and J.~Cort{\'e}s, ``Distributed continuous-time convex
  optimization on weight-balanced digraphs,'' \emph{IEEE Transactions on
  Automatic Control}, vol.~59, no.~3, pp. 781--786, 2014.

\bibitem{tsianos2012push}
K.~I. Tsianos, S.~Lawlor, and M.~G. Rabbat, ``Push-sum distributed dual
  averaging for convex optimization,'' in \emph{Proceedings of the 2012 IEEE
  51st Annual Conference on Decision and Control (CDC)}, Maui, HI, 2012, pp.
  5453--5458.

\bibitem{tsianos2011distributed}
K.~I. Tsianos and M.~G. Rabbat, ``Distributed consensus and optimization under
  communication delays,'' in \emph{Proceedings of the 49th Annual Allerton
  Conference on Communication, Control, and Computing (Allerton), 2011},
  Monticello, IL, 2011, pp. 974--982.

\bibitem{nedic2010constrained}
A.~Nedi{\'c}, A.~Ozdaglar, and P.~A. Parrilo, ``Constrained consensus and
  optimization in multi-agent networks,'' \emph{IEEE Transactions on Automatic
  Control}, vol.~55, no.~4, pp. 922--938, 2010.

\bibitem{zhu2013approximate}
M.~Zhu and S.~Mart{\'\i}nez, ``An approximate dual subgradient algorithm for
  multi-agent non-convex optimization,'' \emph{IEEE Transactions on Automatic
  Control}, vol.~58, no.~6, pp. 1534--1539, 2013.

\bibitem{bianchi2011convergence}
P.~Bianchi and J.~Jakubowicz, ``Convergence of a multi-agent projected
  stochastic gradient algorithm for non-convex optimization,'' \emph{IEEE
  Transactions on Automatic Control}, vol.~58, no.~2, pp. 391--405, Feb 2013.

\bibitem{Lorenzo2016NEXT}
P.~D. Lorenzo and G.~Scutari, ``{NEXT:} in-network nonconvex optimization,''
  \emph{IEEE Transactions on Signal and Information Processing over Networks},
  vol.~2, no.~2, pp. 120--136, June 2016.

\bibitem{tatarenko2015non}
T.~Tatarenko and B.~Touri, ``Non-convex distributed optimization,'' \emph{arXiv
  preprint arXiv:1512.00895}, Dec. 2015.

\bibitem{gharesifard2010does}
B.~Gharesifard and J.~Cort{\'e}s, ``When does a digraph admit a doubly
  stochastic adjacency matrix?'' in \emph{Proceedings of the 2010 American
  Control Conference}, Baltimore, MD, June 2010, pp. 2440--2445.

\bibitem{ScuFacSonPalPan2014}
G.~Scutari, F.~Facchinei, P.~Song, D.~P. Palomar, and J.-S. Pang,
  ``Decomposition by partial linearization: Parallel optimization of
  multi-agent systems,'' \emph{IEEE Transactions on Signal Processing},
  vol.~62, no.~3, pp. 641--656, Feb. 2014.

\bibitem{facchinei2015parallel}
F.~Facchinei, G.~Scutari, and S.~Sagratella, ``Parallel selective algorithms
  for nonconvex big data optimization,'' \emph{IEEE Transactions on Signal
  Processing}, vol.~63, no.~7, pp. 1874--1889, 2015.

\bibitem{kempe2003gossip}
D.~Kempe, A.~Dobra, and J.~Gehrke, ``Gossip-based computation of aggregate
  information,'' in \emph{Proceedings of the 44th Annual IEEE Symposium on
  Foundations of Computer Science, 2003}, Oct. 2003, pp. 482--491.

\bibitem{nedic2015distributed}
A.~Nedic and A.~Olshevsky, ``Distributed optimization over time-varying
  directed graphs,'' \emph{IEEE Transactions on Automatic Control}, vol.~60,
  no.~3, pp. 601--615, 2015.

\bibitem{UnboundedGradient}
Y.~Sun and G.~Scutari, ``Distributed nonconvex optimization for sparse
  representation,'' Purdue University, Tech. Rep., May 2016.

\bibitem{bertsekas1999nonlinear}
D.~P. Bertsekas, \emph{Nonlinear programming}.\hskip 1em plus 0.5em minus
  0.4em\relax Athena Scientific, 2 ed., 1999.

\bibitem{blondel2005convergence}
V.~D. Blondel, J.~M. Hendrickx, A.~Olshevsky, and J.~N. Tsitsiklis,
  ``Convergence in multiagent coordination, consensus, and flocking,'' in
  \emph{Proceedings of the 44th IEEE Conference on Decision and Control},
  Seville, Spain, Dec. 2005, pp. 2996--3000.

\bibitem{scherber2004locally}
D.~S. Scherber and H.~C. Papadopoulos, ``Locally constructed algorithms for
  distributed computations in ad-hoc networks,'' in \emph{Proceedings of the
  3rd international symposium on Information processing in sensor
  networks}.\hskip 1em plus 0.5em minus 0.4em\relax Berkeley, CA: ACM, 2004,
  pp. 11--19.

\bibitem{xiao2005scheme}
L.~Xiao, S.~Boyd, and S.~Lall, ``A scheme for robust distributed sensor fusion
  based on average consensus,'' in \emph{International Symposium on Information
  Processing in Sensor Networks, 2005.}, Los Angeles, CA, April 2005, pp.
  63--70.

\bibitem{Lorenzo2015NEXT-CAMSAP15}
P.~D. Lorenzo and G.~Scutari, ``Distributed nonconvex optimization over
  networks,'' in \emph{Proceedings of the IEEE International Conference on
  Computational Advances in Multi-Sensor Adaptive Processing (CAMSAP 2015)},
  Dec. 13--16 2015.

\bibitem{Lorenzo2015NEXT-ICASSP16}
------, ``Distributed nonconvex optimization over time-varying networks,'' in
  \emph{Proceedings of the IEEE International Conference on Acoustics, Speech,
  and Signal Processing (ICASSP 16)}, March 20--25 2016.

\bibitem{Xu2015augmented}
J.~Xu, S.~Zhu, Y.~C. Soh, and L.~Xie, ``Augmented distributed gradient methods
  for multi-agent optimization under uncoordinated constant stepsizes,'' in
  \emph{Proceedings of the 54th IEEE Conference on Decision and Control (CDC)},
  Dec 2015, pp. 2055--2060.

\bibitem{nedich2016achieving}
A.~Nedich, A.~Olshevsky, and W.~Shi, ``Achieving geometric convergence for
  distributed optimization over time-varying graphs,'' \emph{arXiv preprint
  arXiv:1607.03218}, Jul. 2016.

\bibitem{XiKha-J'16}
C.~Xi and U.~A. Khan, ``{ADD-OPT: A}ccelerated distributed directed
  optimization,'' \emph{arXiv preprint arXiv:1607.04757}, Jul. 2016.

\bibitem{qu2016harnessing}
G.~Qu and N.~Li, ``Harnessing smoothness to accelerate distributed
  optimization,'' \emph{arXiv preprint arXiv:1605.07112}, May 2016.

\end{thebibliography}

\end{document}